\documentclass{PoS}
\def\Journal#1#2#3#4{{#1} {\bf #2}, #3 (#4)}

\def\be{\begin{equation}}
\def\ee{\end{equation}}
\def\bea{\begin{eqnarray}}
\def\eea{\end{eqnarray}}

\def\eros{{\sc EROS}}
\def\eros2{{\sc EROS}{\rm -2}}
\def\eros1{{\sc EROS}{\rm -1}}
\def\macho{{\sc MACHO}}
\def\supermacho{{\sc Super-MACHO}}
\def\ogle{{\sc OGLE}}

\def\moa{{\sc MOA}}

\def\lmc{{\sc LMC}}
\def\smc{{\sc SMC}}

\def\e1cl{{\sc eros}1-{\sc lmc}-}

\def\eros{{\sc EROS}}
\def\macho{{\sc MACHO }}
\def\lmc{{\sc LMC}}
\def\smc{{\sc SMC}}

\def\tempest%
{\begin{array}{ccc}
1 & 1 & 1 \\
1 & 1 & 1 \\
4 & 3 & 8
\end{array}}

\def\e{{\rm E}}

\def\Msol{M_\odot} 
\def\deg{\rm deg}
\def\degr{^\circ}
\def\day{\rm d}
\def\te{t_{\rm E}}

\title{Review of results from the EROS microlensing search for
massive compact objects}

\ShortTitle{Review of results from EROS}

\author{\speaker{M.~Moniez}\thanks{on behalf of the EROS collaboration.}\\
        Laboratoire de l'Acc\'{e}l\'{e}rateur Lin\'{e}aire,
IN2P3 CNRS, Universit\'e Paris-Sud, 91405 Orsay Cedex, France\\
        E-mail: \email{moniez@lal.in2p3.fr}}

\abstract{
We present the results of the
\eros2 search for the hidden galactic matter of the halo through
the gravitational microlensing of stars in the Magellanic clouds.
Microlensing was also searched for and found in the Milky-Way
plane, where foreground faint stars are expected to lens background
stars.
A total of 67 million of stars
were monitored over a period of about 7 years.
Hundreds of microlensing candidates have been found in the
galactic plane, but only one
was found towards the subsample of bright --well measured--
Magellanic stars. This result implies that
massive compact halo objects (machos) in the
mass range $10^{-7}M_\odot<M<5M_{\odot}$ are ruled out
as a major component of the Milky Way Halo.
	}

\FullConference{Identification of dark matter 2008\\
		 August 18-22, 2008\\
		 Stockholm, Sweden}

\begin{document}

\section{Introduction} 
From 1990 to 2003, EROS team has performed a large program
of microlensing survey towards the Magellanic Clouds (LMC and SMC),
the Galactic center (CG) and the Galactic spiral arms (GSA),
as far as $55\degr$ longitude away from the galactic center.
Gravitational microlensing~\cite{Pac1986} occurs if a massive
compact object passes close enough to the line of sight of a star,
temporarily magnifying its light.
In the approximation of a single point-like object moving
with a relative constant speed in front of
a single point-like source,
the visible result is an achromatic and symmetric variation of
the apparent source luminosity as a function of time.
The ``lensing time scale" $t_E$, given by the ratio between the
Einstein radius of the deflector and its transverse speed $V_T$,
is the only measurable parameter bringing useful information on the lens
configuration:
$$
t_E (days)=78\times
\left[\frac{V_T}{100 km/s}\right]^{-1}
\times \left[\frac{M}{\Msol}\right]^{\frac{1}{2}}
\times \left[\frac{L}{10\ Kpc}\right]^{\frac{1}{2}}
\times\frac{[x(1-x)]^{\frac{1}{2}}}{0.5}\; ,
$$
where $L$ is the distance to the source, $xL$ is the distance to
the deflector and $M$ its mass.
The optical depth $\tau$ towards a target is defined as the
average probability for the line of sight of a target
star to intercept the Einstein ring
of a deflector (producing a magnification $> 1.34$).
\section{Observations and data reduction}
\eros2\ uses a 
$\sim 1\; \deg^2$ CCD mosaic mounted on the MARLY 1 meter
diameter telescope installed at the La Silla ESO observatory
(see~\cite{bau97} and references therein).
We monitored $98\ \deg^2$ in the Magellanic Clouds,
$60\ \deg^2$ in the CG, and $29\ \deg^2$ in the GSA.
Images were taken simultaneously in two wide passbands.
Each field has been measured a few hundred times in each passband.
The production of light curves proceeded in three steps :
template images construction, star catalog production from the templates, 
and photometry of individual images to obtain the light curves.
Our catalogues contain
about $29.2\times10^6$ objects from the LMC,
$4.2\times10^6$ from the SMC,
$20.\times10^7$ from the CG and
$12.9\times10^7$ from the GSA.
After alignment with the catalogue,
photometry is performed on each image
with soft\-ware spe\-cific\-ally designed for
crowded fields, PEIDA (Photo\-m\'etrie et \'Etude d'Images Destin\'ees
\`a l'Astrophysique)~\cite{ANS96}.
\section{Using the brightest stars to escape the blending problems}
Using sophisticated simulations, we found that the optical depth
underestimate due to the microlensing magnification underestimate
induced by source confusion (blending) is compensated
by extra events due to faint stars within the seeing disk
of resolved objects.
Nevertheless, considering the size of the effects, we decided
to consider only the subsample of the brightest stars,
that do much less suffer from blending complications,
to obtain reliable microlensing optical depth estimates towards the 
Magellanic clouds and the Galactic center.
We then concentrated on the clump red-giant stars towards
the CG, and on the stars with $R_{eros}<18.2$ to $19.7$
(depending on the field density) towards the LMC.
Another advantage to use these brightest stars is that they also
benefit from the best photometric resolution.
The philosophy is somewhat different towards the Galactic spiral arms,
because of, contrary to the other targets,
the distance of the sources is widely distributed and poorly known.
We therefore decided to use all the stars for the optical depth
estimates, and to define the concept of ``catalogue optical depth'',
that is relative to our specific catalogue of monitored stars.
The interpretation of this optical depth requires
a careful modelling of the galaxy plane as it results from an
average of optical depth on a continuum of source distances.
The final sample of light-curves on which we have searched
for microlensing then contains respectively
$6.05\times10^6$ and $0.9\times10^6$ bright stars towards LMC
and SMC,
$5.6\times10^7$ clump-giant stars towards the CG and
$12.9\times10^7$ stars towards the GSA.
\section{The search for lensed stars}
The general philosophy for the event selection is common to all
the targets.
Details on the
analysis of CG, LMC and GSA can be found in~\cite{hamadache},
~\cite{tisserand} and~\cite{rahal}.

We first searched for bumps in the light curves, that
we characterized by their probability
to be due to accidental occurrence on a stable star light curve.
We select the light curves that have a significant
positive fluctuation in both colors with a sufficient
time overlap.
To reject most of the periodic or irregular variable stars, we
remove the light curves that have significant other bumps
({\it positive or negative}).
After this filtering, the remaining light curves can be fitted for the
microlensing hypothesis, and the final selection is based on variables
using the fitted parameters.
We apply criteria  
devised so as to select microlensing events, keeping in mind that such
analysis should also detect events
with second order effects such as parallax, binary lens...
Specific rejection criteria against background supernovae
have also been applied towards the Magellanic clouds.
We estimate our detection efficiency
using the technique of the superposition of simulated events on
experimental light curves from an unbiased sub-sample of our catalogues.
\section{Candidate sample}
\begin{figure}
\centering
\parbox{8.5cm}{
\includegraphics[width=8.cm]{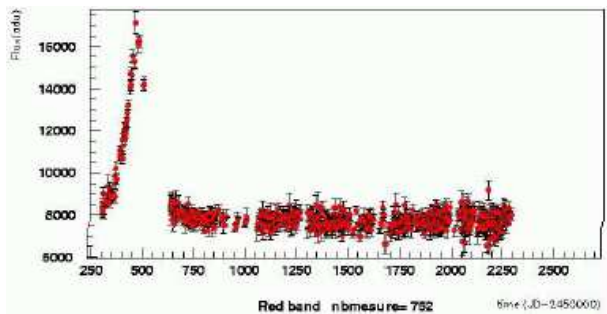}
}
\parbox{5cm}{\caption[] 
{\it
Light-curve of the SMC microlensing event
in the EROS red-passband.
}
\label{cand-SMC}}
\end{figure}

We found 120 events towards the sample of clump-giants of the
CG~\cite{hamadache}, 26 events towards the GSA~\cite{rahal},
and only one event~\cite{EROSMC} (see Fig. \ref{cand-SMC})
towards the bright stellar population of SMC/LMC.
With respect to previous EROS publications~\cite{erosplates,las00},
the number of events
towards LMC/SMC has changed. Amongst the reasons are the fact that
we now concentrate on the bright stars, and the fact that candidates
died because they exhibited another significant bump years after their
selection.
\section{Discussion. Limits on the abundance of machos}
We find that the optical depths towards the CG and
the 4 targets in the GSA are in
good agreement with the predictions from the galactic
models~\cite{hamadache,GSA3y} (Fig. 2).
\begin{figure}[h]
\centering
\parbox{7.cm}{
\includegraphics[width=6.cm]{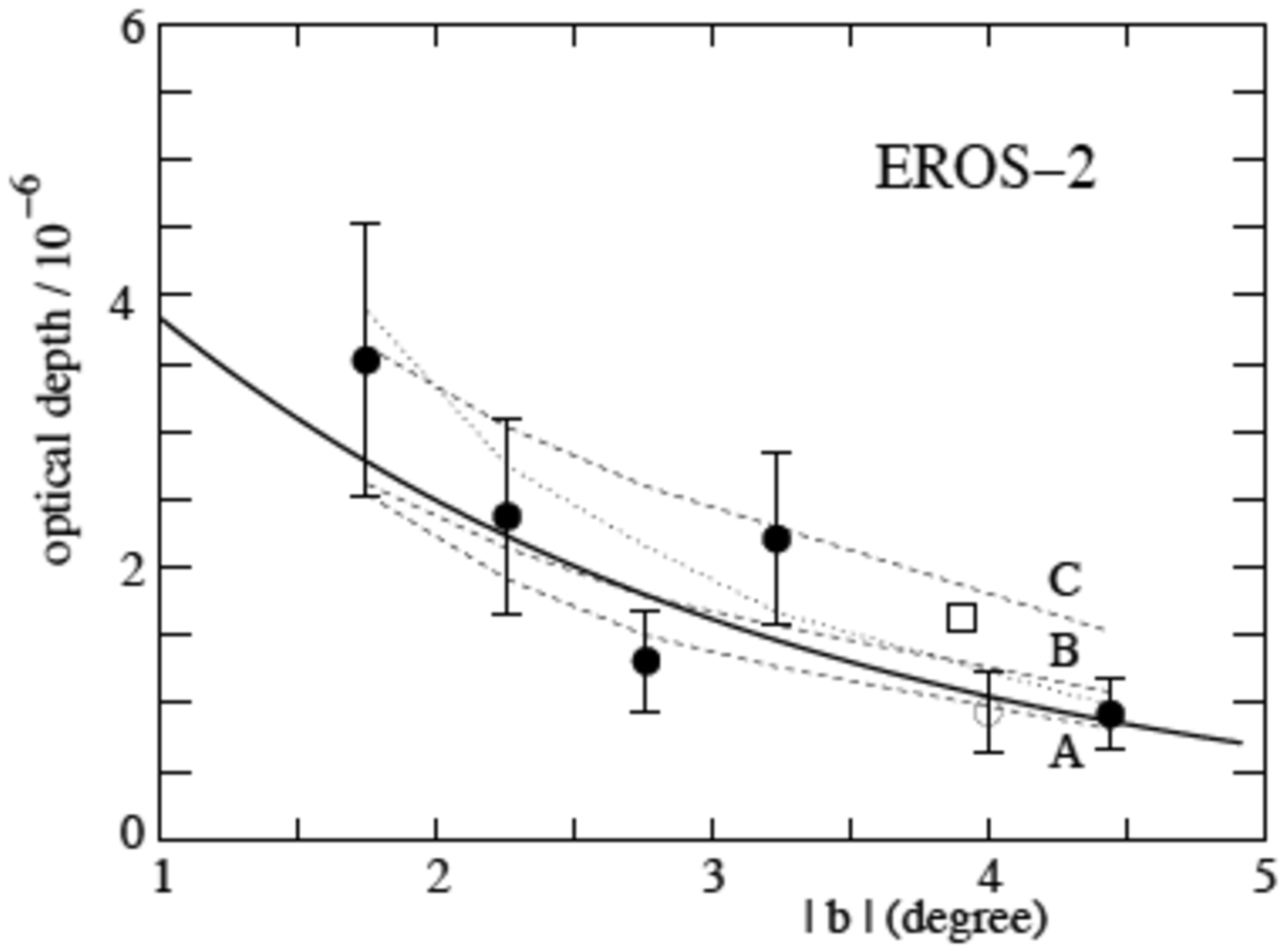}
\caption[] 
{\it
Measured and expected
optical depth as a function of the galactic latitude (up) and
of the galactic longitude (right).
}
}
\parbox{7cm}{
\includegraphics[width=7.cm,height=7.cm]{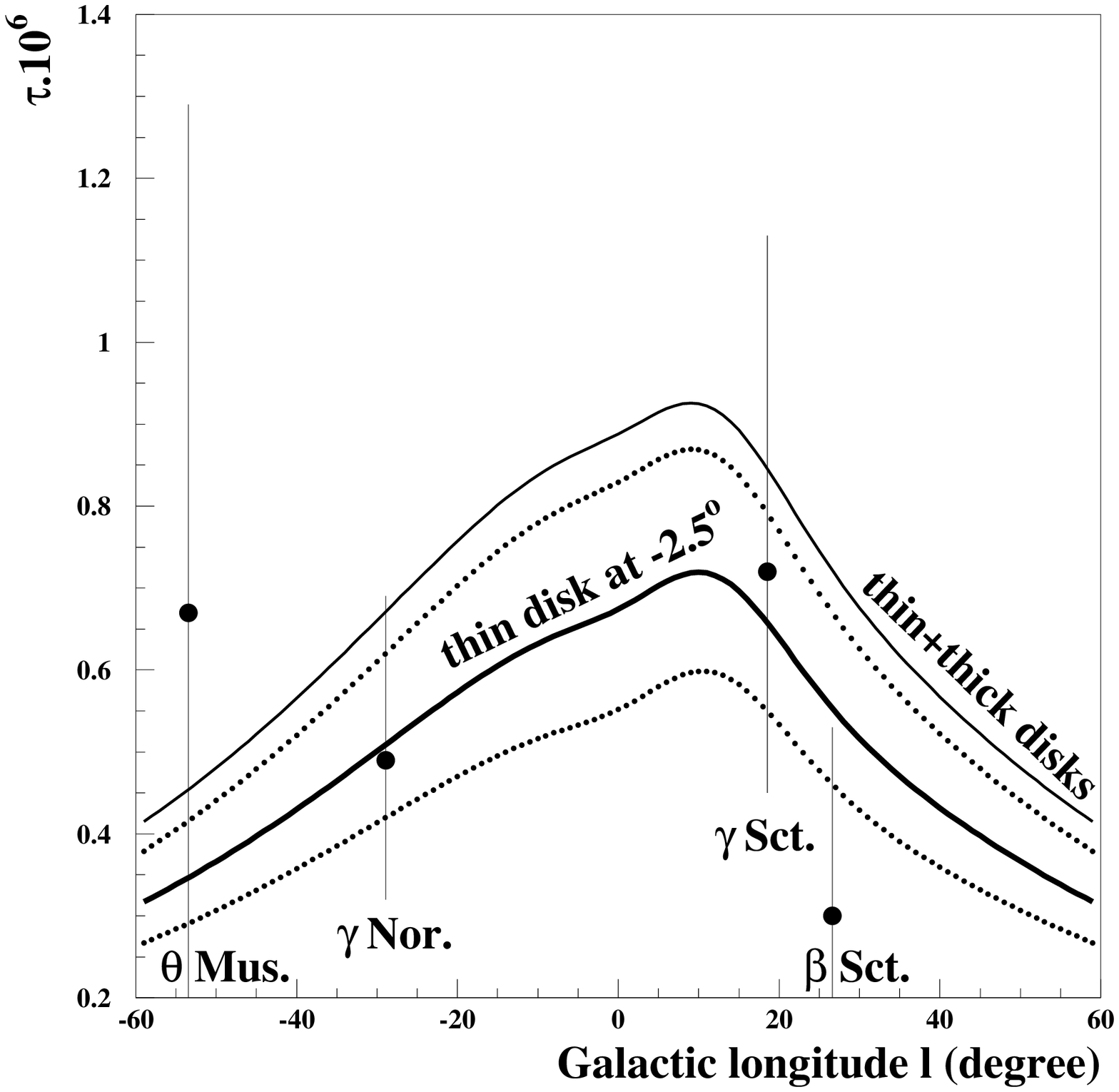}
}
\label{tauCG}
\end{figure}

\begin{figure}
\centering
\parbox{8.5cm}{
\includegraphics[width=8cm,height=5cm]{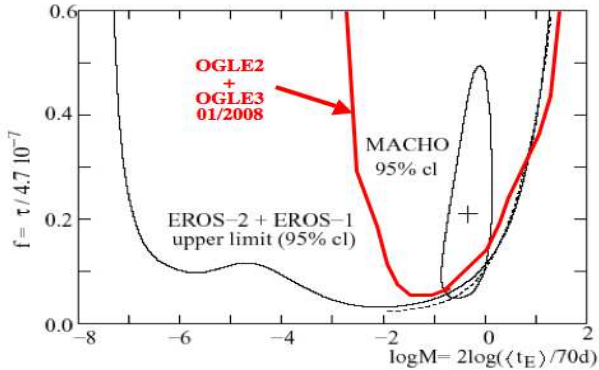}
}
\parbox{5cm}{\caption[] 
{\it
The solid line shows the \eros\ upper
limit on the contribution of compact objects to a standard
spherical Galactic halo, as a function
of their mass, based on zero observed \lmc\ events and
assuming that the one observed \smc\ event is not due to halo lensing.
}    
\label{newlim}}
\end{figure}

In contrast,
we have found only one event towards the Magellanic clouds,
whereas $\sim50$ events would have been expected if
the halo were entirely populated by objects of
mass $0.2M_{\odot}<M<0.8M_{\odot}$.
We then deduce an upper limit on the contribution of the
compact objects to the so-called standard spherical halo
(see Fig. \ref{newlim}).
This limit can also be expressed in optical depth.
In the $\te$ range favored by  the \macho\ collaboration, we 
find
$\tau_{\rm lmc} <\;0.3 \times 10^{-7}\; at \; 95\%\; CL$,
in clear conflict with the value of the \macho\ collaboration,
$\tau_{\rm lmc}=1.2^{+0.4}_{-0.3}\times10^{-7}$,
based on the observation of 17 events~\cite{MachoLMC},
but in excellent agreement with the recently published
results from the OGLE collaboration~\cite{OGLEres}.
For the SMC, the one observed event corresponds to an optical depth of 
$1.7\times 10^{-7}$.
Taking into account
only Poisson statistics on one event, this gives
$.085\times10^{-7} < \tau_{\rm smc} < 8.7\times 10^{-7} \; at\; 95\%\; CL$.
This is consistent with the expectations of lensing by objects in the 
SMC itself~\cite{graffsmc},
$\tau_{\rm smc-smc}\sim0.4\times10^{-7}$.
The value of $\te=120\; days$ is also consistent with expectations for 
self-lensing as $\langle\te\rangle\sim 100\; days$
for a mean lens mass of $0.35M_{\odot}$. We also note that 
the self-lensing interpretation is favored from the absence of an 
indication of parallax in the light curve~\cite{parsmc}.

There are considerable differences between the EROS and MACHO
data sets that may explain the conflict.
Generally speaking, MACHO uses faint stars in dense fields
($1.2\times 10^7$ stars over $14\;\deg^2$) while \eros2\ uses bright stars in
sparse fields ($0.7\times 10^7$ stars over $90\;\deg^2$).
As a consequence of the use of faint stars, only two of the 17 MACHO
candidates are sufficiently bright to be compared to our bright sample
(and the corresponding events occurred before EROS data taking).
The use of dense fields by the MACHO group also suggests that the
higher MACHO optical depth may be due, in part, to self lensing in
the inner parts of the LMC.
The contamination of irregular variable objects faking microlensing in
low photometric resolution events should also be stronger in
the faint sample of stars used by MACHO.
As already mentioned, another problem
with the use of faint source stars is 
the large blending effects that must be understood.
The experience
with the use of faint stars in the Galactic Bulge suggest that
the uncertainty induced by the blending effects in such a
sample may be underestimated.
\section{Conclusions}
The \eros\ and \macho\ programs were primarily motivated by the
search for halo brown dwarfs of $M\sim0.07M_\odot$.
EROS has demonstrated its sensitivity to microlensing events,
finding microlensing rates compatible with the predictions of the
galactic models in the Galactic plane.
The lack of events towards the Magellanic clouds clearly indicates
that the Galactic hidden matter is not made of compact objects in the
mass range $10^{-7}M_\odot<M<5M_{\odot}$.
Whatever the source of the disagreement between EROS and MACHO on
this subject, we can hope that
new data from the \ogle3\ , \moa\ and \supermacho\ collaborations
will settle the matter.

\end{document}